\documentclass[aps,pra,twocolumn,showpacs,groupedaddress,superscriptaddress,10pt]{revtex4-1}
\usepackage{bm,graphicx,amsmath}
\usepackage{placeins}
\usepackage{amssymb}
\usepackage{amsmath}
\usepackage{epsfig}
\usepackage{natbib}
\usepackage{color}
\usepackage[pdftex,colorlinks,linkcolor=blue,citecolor=blue,urlcolor=blue]{hyperref}
\usepackage{subfigure}
\usepackage{array}
\usepackage{multirow} 
\usepackage{booktabs}
\usepackage[normalem]{ulem}
\usepackage{color}
\usepackage{soul}

\newcommand{\comment}[1]{}
\DeclareMathOperator{\sech}{sech}
\definecolor{gray}{gray}{0.6}

\begin{document}
\title{Soliton-based matter wave interferometer}
\author{J. Polo}
\affiliation{Departament de F\'{\i}sica, Universitat Aut\`{o}noma de Barcelona, E-08193 Bellaterra, Spain} 
\email[]{Juan.Polo@uab.cat}
\author{V. Ahufinger}
\affiliation{Departament de F\'{\i}sica, Universitat Aut\`{o}noma de Barcelona, E-08193 Bellaterra, Spain} 
\date{\today}
\begin{abstract}
We consider a matter wave bright soliton interferometer composed of a harmonic potential trap with a Rosen--Morse barrier at its center on which an incident soliton collides and splits into two solitons. These two solitons recombine after a dipole oscillation in the trap at the position of the barrier. We focus on the characterization of the splitting process in the case in which the reflected and transmitted solitons have the same number of atoms. We obtain that the velocity of the split solitons strongly depends on the nonlinearity and on the width of the barrier and that the reflected soliton is in general slower than the transmitted one. Also, we study the phase difference acquired between the two solitons during the splitting and we fit semi-analytically the main dependences with the velocity of the incident soliton, the nonlinearity and the width of the barrier. The implementation of the full interferometer sequence is tested by means of the phase imprinting method.
\end{abstract}
\maketitle


\section{Introduction}
\label{sec:Introduction}

The inherent atomic properties, like mass or polarizability, and the very high sensitivities achievable with atom interferometers make them a very versatile tool for high-precision measurements of, for instance, fundamental constants, internal forces, accelerations or rotations \cite{Cronin_2009,Bouchendira_2011,Geiger_2011,Stockton_2011}, being also used in general relativity tests \cite{Dimopoulos_2007,Hohensee_2011} and having even being proposed for gravitational wave detection \cite{Dimopoulos_2008,Graham_2013}.

Atomic Bose-Einstein condensates (BECs) are promising candidates to increase the phase sensitivity of atom interferometers due to their large coherence length \cite{Andrews_1997,Hagley_1999,Bloch_2000}. However, elastic collisions in BECs can produce phase diffusion reducing the phase coherence \cite{Lewenstein_1996,Javanainen_1997,Castin_1997}. Phase diffusion can be reduced for example by using Feshbach resonances \cite{Fattori_2008,Cheng_2010} or by taking advantage of the interactions to introduce non-classical correlations between the two arms of the interferometer \cite{Jo_2007,Berrada_2013}. Nonlinear interactions also give rise to squeezed states which allow to surpass the standard quantum limit \cite{Kitagawa_1993,Esteve_2008,Riedel_2010,Gross_2010,Gross_2011,Lucke_2011,Hamley_2012}. 

In BECs with attractive interactions, the use of matter wave bright solitons \cite{Khaykovich_2002,Hulet_2002,Wieman_2006} for interferometry was already proposed in \cite{Hulet_2002} and its potential to increase phase sensitivity has been recently discussed  \cite{Kasevich_2012}. The main advantages that matter wave bright solitons offer are that they can be described by single large mass macroscopic wavefunctions, have a well defined spatial localization and present absence of dispersion. Some methods have been proposed for implementing the beam splitter behavior required in a matter wave bright soliton interferometer \cite{Malomed_2013} such as applying a resonant $\pi/2$ pulse to an internal state transition of the soliton \cite{Gardiner_2011}, using an accurate control of the scattering length in space or in time to split the soliton into two or more pieces \cite{Paredes_2012} or collisions with a potential barrier in different scenarios like using a rectangular barrier \cite{Carr_2012,Damgaard_2012,Sakaguchi_2005}, Gaussian and delta type potential barriers \cite{Gardiner_2012,Holmer_2007,Holmer_20071,Gertjerenken_2012,Gertjerenken_2012b,Weiss_2012}, in a quasi one dimensional external harmonic confinement in the presence of quantum fluctuations \cite{Martin_2012} or considering three dimensional dynamics \cite{Cuevas_2013}. 

Here, we consider a matter wave bright soliton interferometer composed of a harmonic potential trap with a Rosen--Morse barrier at its center on which an incident soliton collides and splits into two solitons. The two split solitons recombine after a dipole oscillation in the trap at the position of the barrier producing two output solitons.
The number of atoms of these two outputs provides a measure of the phase difference between the two arms of the interferometer. The phase difference acquired by the two solitons during the splitting process in a collision with a potential barrier is often assumed to be $\pi/2$ even for finite width barriers. Here, we show that this is only the case in the limit of very high velocities of the incident soliton and extremely narrow barriers. In general, the phase difference between the split solitons strongly depends on the velocity of the incident soliton, the nonlinearity and the width of the barrier. We also point out the limitations to achieve a symmetric splitting of the incident soliton by scattering on a potential barrier. Although the two split solitons can be obtained with the same number of atoms, in general, the reflected soliton has less velocity than the transmitted one. 

The paper is organized as follows. In section \ref{sec:Physical system} we describe the considered matter wave bright soliton interferometer. In section \ref{sec:Transmission coefficient} we study the transmission coefficient as a function of the kinetic energy of the soliton for different nonlinearities. Section \ref{sec:Splitting} is devoted to the analysis of the splitting process focusing on the case of equal-sized splitting. First, in section \ref{sec:RMvsDelta} the area of the Rosen--Morse barrier necessary to obtain two split solitons with the same number of atoms is analyzed. Then, in section \ref{Velocity of the split solitons}, we study the velocity of the split solitons and finally, in section \ref{sec:phase difference} the phase difference acquired between the two split solitons is characterized. Section \ref{sec:Recombination} is dedicated to the recombination process, and finally in section \ref{conclusions} we present the conclusions.

\section{Physical system}
\label{sec:Physical system}

Within the mean field approach the dynamics of a Bose-Einstein condensate at zero temperature in one dimension (1D) is described by the time-dependent 1D Gross--Pitaevskii equation (GPE):
\begin{equation}
\label{GP}
i\hbar\frac{\partial}{\partial t}\Psi({z},t)=\left(-\frac{\hbar^2}{2m}\frac{\partial^2}{\partial z^2}+V({z})+g_{1D}|\Psi({z},t)|^2\right)\Psi({z},t),
\end{equation}
where $V(z)$ is the external potential, $m$ the atomic mass and the parameter that determines the strength of the atom--atom interactions is $g_{1D}=2N\hbar\omega_r a_s$; with $N$, $\omega_{r}$, $a_{s}$ corresponding to the atom number, frequency of the radial confinement and s-wave scattering length, respectively. The wavefunction is normalized to $1$, and, we consider negative scattering lengths, $a_{s}<0$, corresponding to attractive interactions. 

\begin{figure}[htbp]
	\centering
		\includegraphics[width=0.5\textwidth]{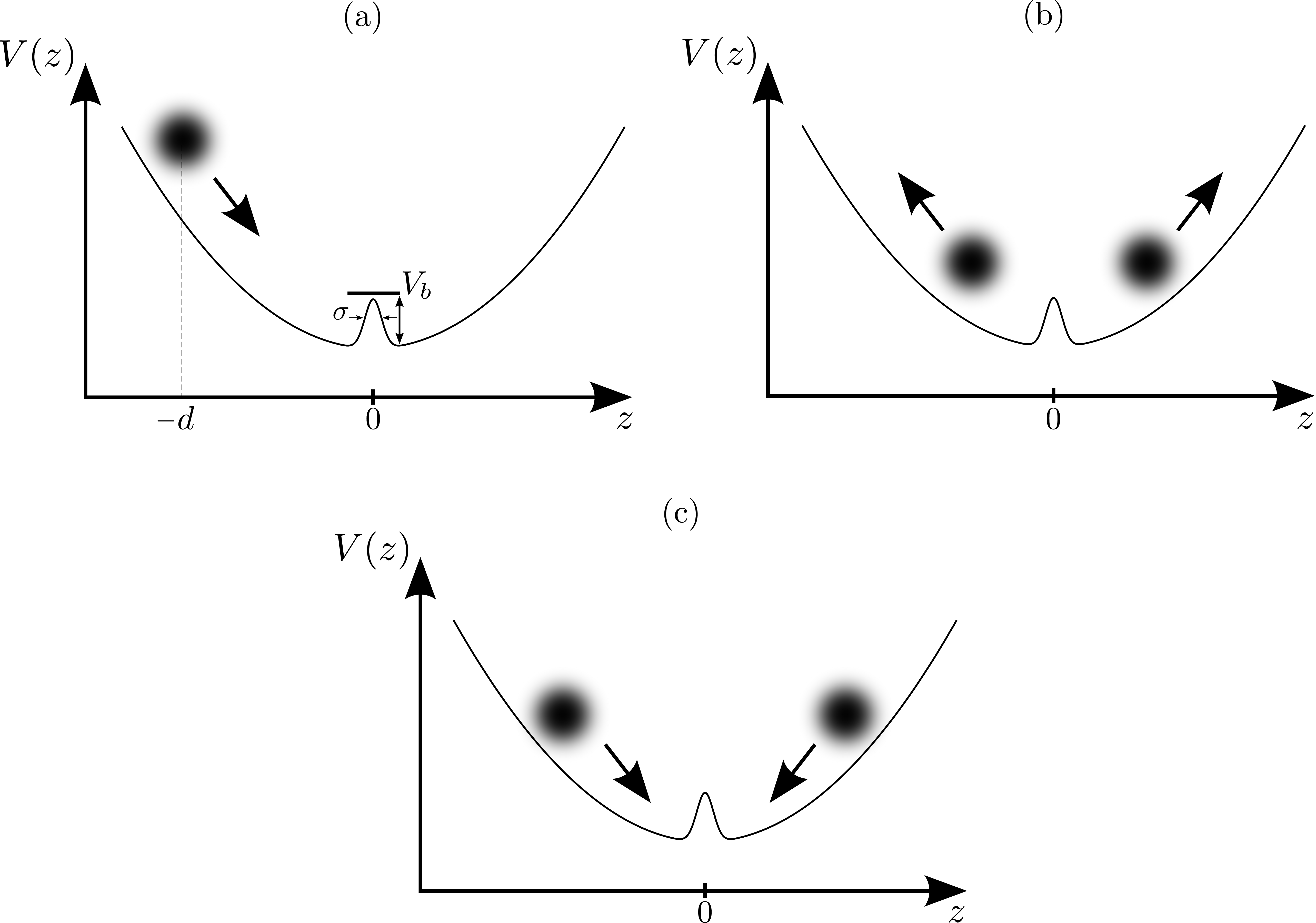}
	\caption{Schematics of the interferometer sequence: (a) initial situation in which the soliton is displaced with respect to the center of the harmonic potential trap by
a distance $d$; (b) the two split solitons obtained after the collision with the barrier separate from each other and (c) the two solitons return to the position of the barrier and collide.}
	\label{fig:scheme}
\end{figure}

For the implementation of the considered interferometer, first a matter wave bright soliton is created in a harmonic external potential trap. Then, the potential trap  is suddenly displaced a distance $d$ in the $z$ direction and the soliton acquires potential energy (Fig. \ref{fig:scheme} (a)). The potential energy given by the displacement, according to the particle models \cite{Poletti_2008,Martin_2007}, is fully converted into kinetic energy once the soliton reaches the center of the trap $E^{v}_{k}= E_{p}=\frac{1}{2}m\omega_{z}^{2}d^{2}$. At this time, a Rosen--Morse potential barrier, on which the soliton will collide, is located at the center of the harmonic potential and then, the external potential reads:
\begin{equation}
\label{external_potential}
V(z)=\frac{1}{2}m\omega_{z}^{2}z^{2} + V_{b}\sech^{2}\left(\frac{z}{\sigma}\right),
\end{equation}
where $\omega_{z}$ is the frequency of the axial confinement, and $V_b$ and $\sigma$ are the strength and the width of the Rosen--Morse potential barrier, respectively. By scattering on the barrier, the incident soliton splits into two solitons which propagate in opposite directions undergoing a dipole oscillation in the harmonic potential (Fig. \ref{fig:scheme} (b)). Finally, the two solitons recombine at the position of the barrier (Fig. \ref{fig:scheme} (c)) producing two output solitons. The number of atoms of these two output solitons provides a measure of the phase difference between the two arms of the interferometer. 

\section{Transmission coefficient}
\label{sec:Transmission coefficient}

The first requirement for the implementation of a matter wave bright soliton interferometer is to possess a mechanism to coherently split the incident soliton in two identical solitons. In  our system, such a mechanism is provided by the interaction with a Rosen--Morse potential barrier, as described in section \ref{sec:Physical system}. The Rosen--Morse potential is a sech-squared-shape potential with analytical solution in the linear regime and provides a good approximation to the potential created by a focused light beam by means of the dipole light force \cite{Lee_2006,Yuri_2011}. Also, the absence of sharp edges in the Rosen--Morse barrier, contrarily to the delta and squared potentials, avoids sharp point effects \cite{Lee_2006}.

\begin{figure}[htbp]
	\centering
		\includegraphics[width=0.50\textwidth]{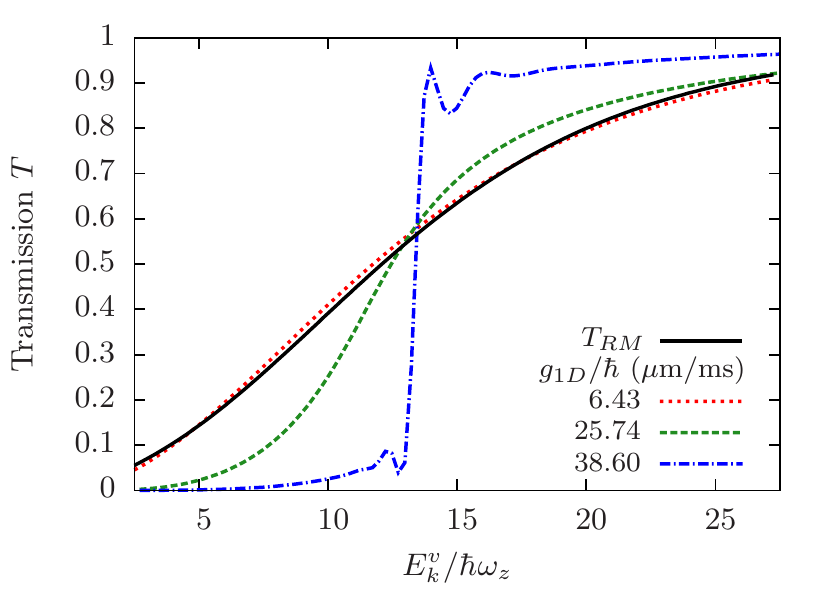}
	\caption{(Color online) Transmission coefficient as a function of the kinetic energy of the incident soliton for different values of the nonlinear interaction term. The solid line corresponds to the analytical solution of the Rosen--Morse potential barrier obtained in the linear regime. The parameter values are: $V_{b}=17.14\hbar\omega_z$, $\sigma=0.67$ $\mu$m and $\omega_{z}=2\pi\times78$ Hz.}
	\label{fig:transmission}
\end{figure}

Fig. \ref{fig:transmission} shows the transmission coefficient as a function of the kinetic energy of the incident soliton, $E^{v}_{k}$, for different values of the nonlinear interaction term. $E^{v}_{k}$ does not contain the quantum pressure term, i.e., it is only due to the gradient of the soliton phase (see Appendix A). The transmission coefficient is defined as:
\begin{equation}
T=\int^{\infty}_{0}{\left|\Psi(z,t=\pi/\omega_{z})\right|^{2}dz},
\end{equation}
and it is obtained numerically at a time such that the two split solitons are well separated from each other ($t=\pi/\omega_{z}$).
From Fig. \ref{fig:transmission}, it can be seen that the nonlinearity dominates the behavior of the transmission coefficient. The kinetic energy of the incident soliton necessary to obtain a fixed value of the transmission coefficient decreases (increases) with the nonlinearity for $T>\overline{T}$ ($T<\overline{T}$), where $\overline{T}$ is a value around transmission coefficients of 0.5, being $\overline{T}=0.57$ for the case shown in Fig. \ref{fig:transmission}. Taking into account that the nonlinearity tends to hold all the atoms together, as $g_{1D}$ increases, the shape of the transmission coefficient becomes sharper, favoring the transmission (reflection) for $T>\overline{T}$ ($T<\overline{T}$). For large enough strength of the nonlinear interactions, the transmission coefficient presents a step-like behavior in which the incident soliton is either completely transmitted or completely reflected. This step-like behavior has been also reported for squared barriers \cite{Sakaguchi_2005,Damgaard_2012}. The analytical linear transmission coefficient of the Rosen--Morse potential, which for $\frac{8m V_{b}\sigma^{2}}{\hbar^{2}}>1$ reads \cite{Landau_1965}:
\begin{equation}
T_{RM}=\frac{\operatorname{sinh}^{2}\left(\sigma\pi\sqrt{\frac{2m E^{v}_{k}}{\hbar^{2}}}\right)}{\operatorname{sech}^{2}\left(\sigma\pi\sqrt{\frac{2m E^{v}_{k}}{\hbar^{2}}}\right)+\operatorname{cosh}^{2}\left(\frac{\pi}{2}\sqrt{\frac{8mV_{b}\sigma^{2}}{\hbar^{2}}-1}\right)}.
\end{equation}
is plotted also in Fig. \ref{fig:transmission} (solid line) showing that it is in good agreement with the numerical results in the limit of low nonlinearity.


\section{Splitting Process}
\label{sec:Splitting}

In this section we focus on the case where the two split solitons have the same number of atoms, i.e., $T=0.5$. For a fixed width of the barrier and a fixed nonlinearity, the potential strength of the barrier, $V_{b}$, is modified to obtain the equal-sized splitting for different velocities of the incident soliton. We consider velocities of the incident soliton and widths of the barrier achievable in current experiments \cite{Hulet} but also we study very high incident velocities and very narrow barriers to approach the limit of the delta potential barrier. In all the cases $E^{v}_{k}<V_{b}$ i.e., the system is in the tunneling regime. 
For the analysis of the splitting mechanism we switch off the external harmonic potential trap in order to take into account only the effects produced by the interaction between the soliton and the barrier and we focus on three main issues: in section \ref{sec:RMvsDelta} we calculate the area of the potential barrier required to obtain the equal-sized splitting; in section \ref{Velocity of the split solitons} we study the difference between the velocity of the transmitted and reflected solitons and its relation with the velocity of the incident soliton; and in section \ref{sec:phase difference} we characterize the phase difference between the transmitted and reflected solitons.

\subsection{Rosen--Morse vs. Delta barrier}
\label{sec:RMvsDelta}
\begin{figure}[htbp]
	\centering
		\includegraphics[width=0.50\textwidth]{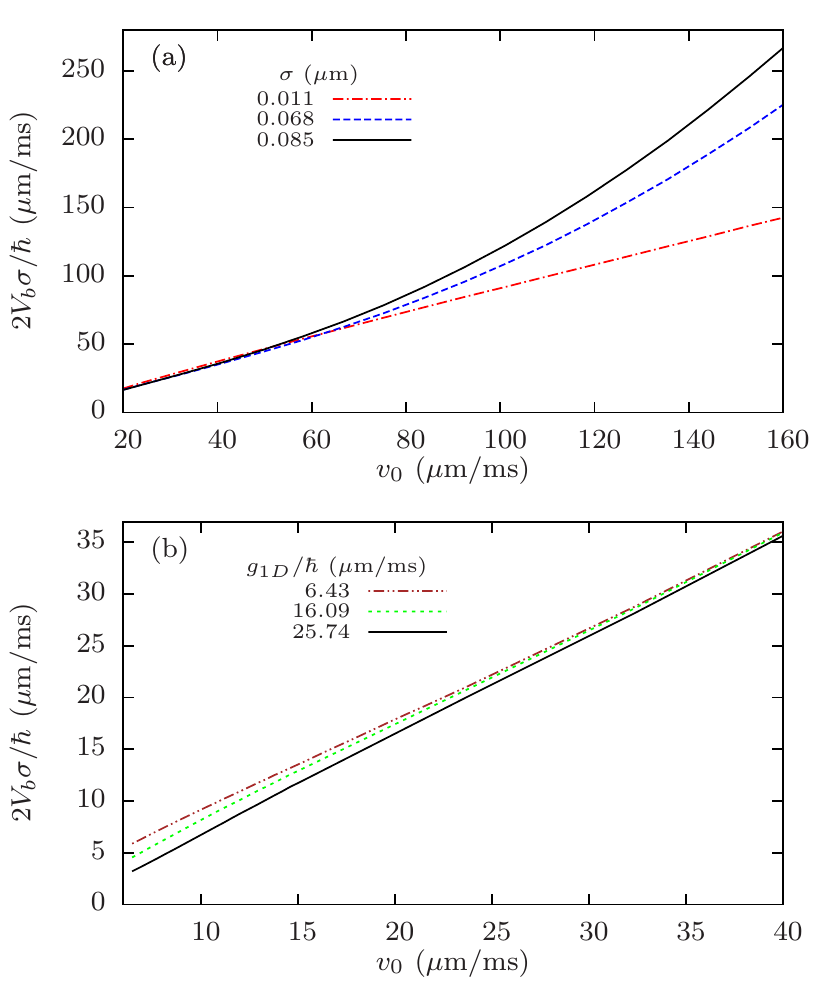}
	\caption{(Color online) Area of the Rosen--Morse barrier as a function of the velocity of the incident soliton, while keeping the transmission coefficient fixed at $T=0.5$, for different values of the width of the barrier and a fixed $g_{1D}/\hbar=25.74$ $\mu$m/ms (a) and for different values of the nonlinear interactions and a fixed width of the barrier $\sigma=0.085$ $\mu$m (b).}
	\label{fig:velo_area}
\end{figure}
In Fig.~\ref{fig:velo_area} we show the area of the Rosen--Morse potential barrier, $A=2V_b\sigma$, required to obtain a fixed transmission coefficient of $0.5$, as a function of the velocity of the incident soliton for different values of the width of the barrier and a fixed nonlinearity (Fig. \ref{fig:velo_area} (a)) and for different values of the nonlinear interaction parameter and a fixed width of the barrier (Fig. \ref{fig:velo_area} (b)). For high velocities of the incident soliton and very narrow barriers (Fig. \ref{fig:velo_area} (a)) the area of the barrier has approximately a linear dependence with the velocity of the incident soliton. Thus, retrieving the behavior for a delta potential barrier, for which the transmission coefficient of a free particle reads \cite{Bohm_1951}:
\begin{equation}
\label{delta}
T_{\bot}=\frac{1}{1+\frac{\lambda^{2}}{\hbar^{2}v_{0}^{2}}},
\end{equation}
with $\lambda$ being the strength of the delta potential barrier, $V_{\bot}=\lambda\delta(z)$, and $v_0$ the velocity of the incident particle. From Eq. (\ref{delta}), in order to keep $T_{\bot}$ equal to $0.5$, the strength of the delta potential barrier must have a linear dependence with the incident velocity $\lambda=\hbar v_0$. In the Rosen--Morse barrier for $T=0.5$, we recover a linear dependence of the area with respect to the incident velocity for $V_b\rightarrow\infty$ and $\sigma\rightarrow0$ while keeping the product $V_b\sigma$ constant, but with a different slope than in the delta potential case. For wider barriers (Fig. \ref{fig:velo_area} (a)), we have found an approximately quadratic behavior of the area of the Rosen--Morse potential barrier with respect to the incident velocity for a fixed width of the barrier and a fixed nonlinearity. We can also see that the growth of the area of the barrier with respect to the velocity of the incident soliton is steeper as the width of the barrier increases. Note that here we have considered very thin barriers because for high incident velocities, the width of the barrier is limited from above since for wide enough barriers the incident soliton splits in more than two pieces.
 In order to analyze the effects of the nonlinearity, in Fig. \ref{fig:velo_area} (b) we study low incident velocities and we observe that as $g_{1D}$ increases, for a fixed width of the barrier, the area of the barrier necessary to keep $T=0.5$ decreases.  This effect is consistent with the dependence on the nonlinearity of the transmission coefficient as a function of $E^{v}_{k}$ (shown in Fig. \ref{fig:transmission}). For a fixed $E_{k}^{v}$ and for $T<\overline{T}=0.57$, as $g_{1D}$ increases, the transmission coefficient decreases. Thus, the potential strength of the barrier should decrease in order to maintain the equal-sized splitting. Note that the effect of the nonlinearity decreases as the incident velocity increases.

\subsection{Velocity of the split solitons}
\label{Velocity of the split solitons}
	\begin{figure}[htbp]
	\centering
		\includegraphics[width=0.50\textwidth]{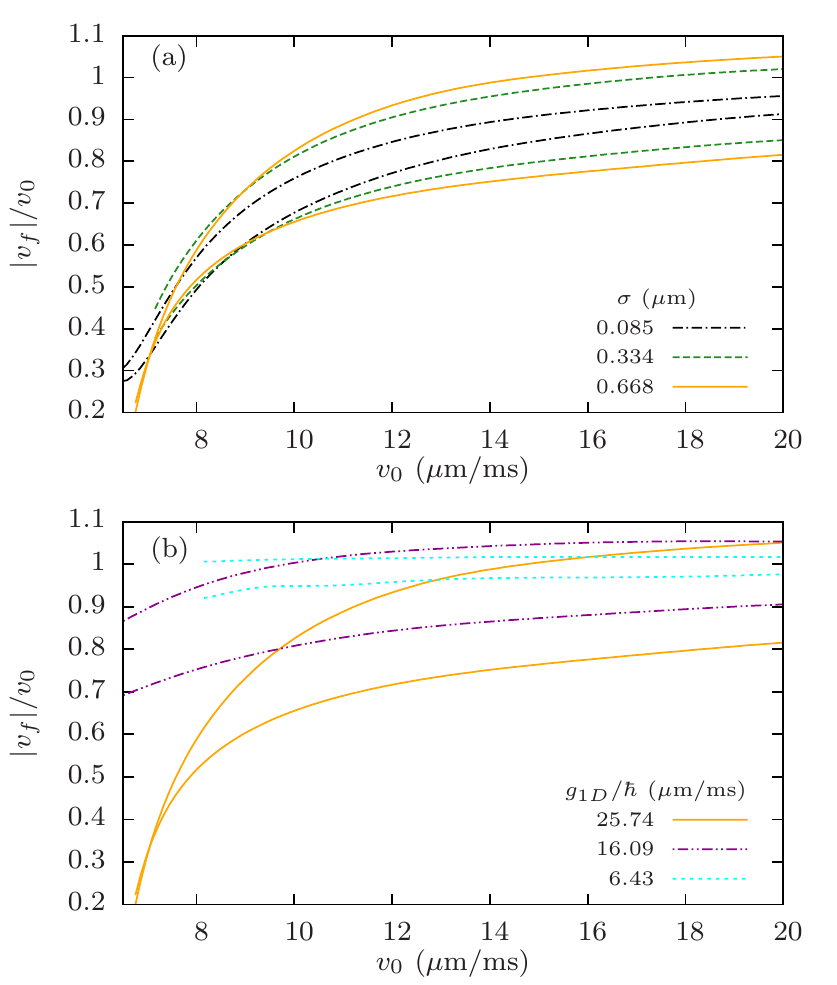}
	\caption{(Color online) Ratio between the modulus of the velocity of the reflected (transmitted) soliton and the velocity of the incident soliton as a function of the velocity of the incident soliton for a fixed $T=0.5$ for different values of the width of the barrier for a fixed $g_{1D}/\hbar=25.74$ $\mu$m/ms (a) and for different values of the nonlinear interactions for a fixed width of the barrier $\sigma=0.668$ $\mu$m (b). The velocities of the transmitted and reflected solitons are represented with the same line type, being in each case the velocity of the reflected soliton the lower curve.}
	\label{fig:velocity}
\end{figure}
In the equal-sized splitting, even though the reflected and transmitted solitons have the same number of atoms, they do not behave symmetrically. In general, we find that the reflected soliton is slower than the transmitted one and their velocities depend on the width of the barrier and on the strength of the nonlinear interaction. Fig. \ref{fig:velocity} shows the numerically calculated ratio between the absolute value of the velocity of the reflected (transmitted) soliton and the velocity of the incident soliton as a function of the velocity of the incident soliton, for different values of the width of the barrier for a fixed $g_{1D}$ (Fig. \ref{fig:velocity} (a)) and for different values of the nonlinearity for a fixed $\sigma$ (Fig. \ref{fig:velocity} (b)). In each case, the lower curve corresponds to the reflected soliton. The difference between the absolute values of the velocities of the transmitted and reflected solitons increases as the width of the barrier increases (Fig. \ref{fig:velocity} (a)). We also observe that, for low velocities of the incident soliton, as the nonlinearity increases (Fig. \ref{fig:velocity} (b)), the split solitons are slowed down and eventually they become trapped at the position of the barrier. This effect also appears in rectangular barriers \cite{Damgaard_2012} and limits the maximum value of the nonlinearity in order to maintain the 50-50 splitting.
 Notice also that the mean of the absolute value of the velocities of the two split solitons, $\overline{\Delta v}=(|v_T|+|v_R|)/2$, is practically independent of the width of the barrier (Fig. \ref{fig:velocity} (a)) but it is strongly affected by the nonlinearity (Fig. \ref{fig:velocity} (b)). Also, $\overline{\Delta v}$ approaches the velocity of the incident soliton for high incident velocities  i.e., the ratio $\overline{\Delta v}/v_0$ tends to one as the incident velocity increases. In Fig. \ref{fig:velocity} we can also see that the velocity of the transmitted soliton is, in some cases, larger than the velocity of the incident one. Nevertheless, the increase of velocity of the transmitted soliton is accompanied by a decrease of the velocity of the reflected soliton and therefore the total energy is conserved. The difference in velocity of the split solitons introduces an accumulated phase difference between the two arms of the interferometer, that will be discussed in section \ref{sec:Recombination}.

\subsection{Phase difference}
\label{sec:phase difference}
Here we analyze the phase difference introduced during the splitting process when the two split solitons have the same number of atoms. Performing a detailed analysis of the phase evolution during the splitting of a soliton colliding with a Rosen--Morse potential barrier, we observe a strong dependence on the width of the barrier, velocity of the incident soliton and nonlinearity. These dependences go beyond the one soliton solution (Eqs. (\ref{soliton}) and (\ref{phase_soliton}) in Appendix A) of the GPE and requires to consider the n-soliton solution obtained by Zakharov and Shabat \cite{ZAKHAROV_1972,Gordon_1983}, which shows that neighbor solitons make their presence felt through phase and position shifts (Eq.(\ref{phaseshift}) in Appendix A). In fact, we find that the phase difference introduced during the splitting of a matter wave bright soliton into two solitons by colliding with a potential barrier arises from two main sources, the interaction between soliton and barrier and the interaction between the reflected and transmitted solitons. The influence of the soliton-soliton interactions when two solitons collide at the position of a potential barrier has been recently discussed \cite{Gardiner_2012}.

\subsubsection{High incident velocities}

	\begin{figure}[htbp]
	\centering
		\includegraphics[width=0.50\textwidth]{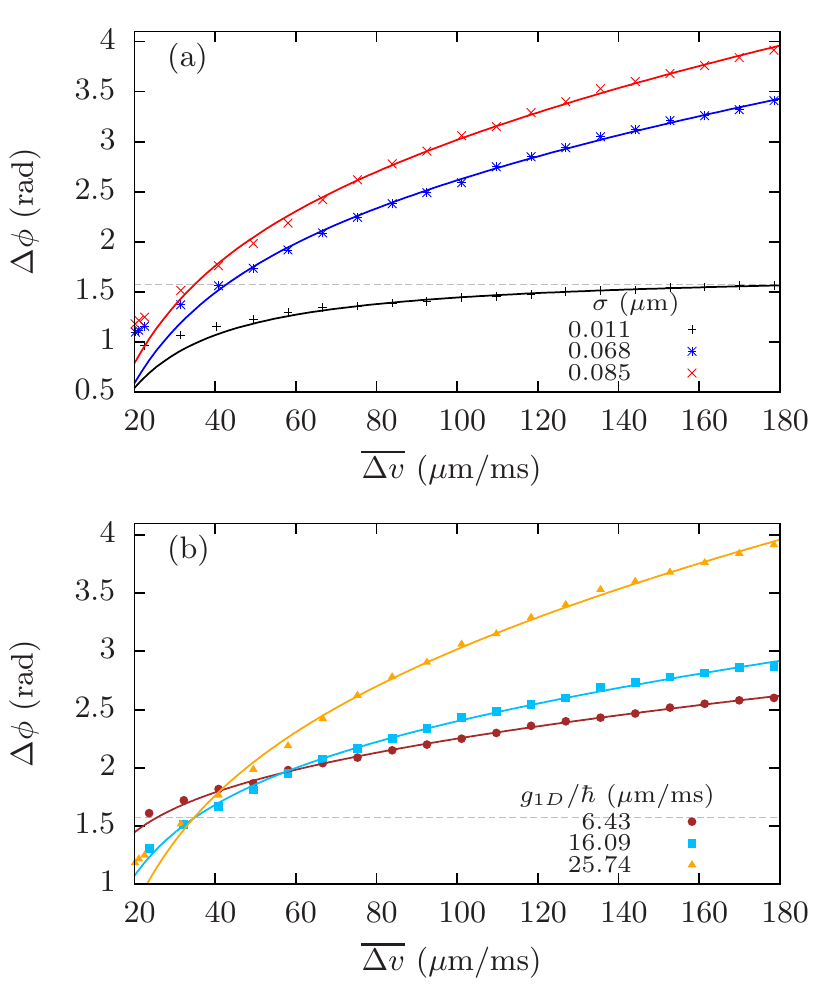}
	\caption{(Color online) Phase difference between the transmitted and reflected solitons as a function of the mean of the absolute values of the velocities of the two split solitons for a fixed transmission coefficient of $0.5$, high velocities of the incident soliton and for different values of the width of the barrier and a fixed $g_{1D}/\hbar=25.74$ $\mu$m/ms (a) and for different values of the nonlinear interactions and a fixed $\sigma=0.085$ $\mu$m (b). Solid lines correspond to the semi-analytical fit given by Eq. (\ref{phase}) with the parameters $C_i$ (i=1,2) adjusted numerically.}
	\label{fig:phase2}
\end{figure}

Fig. \ref{fig:phase2} shows the phase difference between transmitted and reflected solitons for a fixed transmission coefficient, $T=0.5$, as a function of the mean of the absolute values of the velocities of the two split solitons, $\overline{\Delta v}$, for high incident velocities and for different widths of the barrier and a fixed nonlinearity (Fig. \ref{fig:phase2} (a)) and for different nonlinearities and a fixed width of the barrier (Fig. \ref{fig:phase2} (b)). We compute the phase difference of the two split solitons when they are separated $10$ $\mu$m in order to avoid the self-interferences that appear in the reflected soliton just after the splitting. We observe that the phase difference increases as $\overline{\Delta v}$ increases, and its growth depends on the width of the barrier (Fig. \ref{fig:phase2} (a)). For a fixed width of the barrier (Fig. \ref{fig:phase2} (b)), the phase difference introduced during the splitting process in the case of high incident velocities increases as the nonlinearity increases. Taking into account these dependences with the parameters of the system and the phase shift due to the presence of two neighboring solitons, we approximate the phase difference introduced during the splitting of a matter wave bright soliton by interacting with a Rosen--Morse barrier for high velocities of the incident soliton (solid line of Fig. \ref{fig:phase2}) as:

\begin{equation}
\label{phase}
\Delta\phi(\overline{\Delta v})=-2\arctan\left(\frac{g_{1D}}{2\hbar \overline{\Delta v}}\right) + C_1 + C_2\sqrt{\overline{\Delta v}} 
\end{equation}
where $C_i$, with $i=1,2$, are independent of $\overline{\Delta v}$ but depend on the parameters of the system, and in our case have been adjusted numerically.
The first term of the right hand side of Eq. (\ref{phase}) corresponds to the phase shift associated to the soliton-soliton interaction derived from Eq. (\ref{phaseshift}) in Appendix A) and that is highly affected by $g_{1D}$. The second term provides a phase difference due to the interaction of the incident soliton with the barrier that does not depend on $\overline{\Delta v}$ but in general is strongly affected by the nonlinearity, growing as the nonlinearity increases. The third term depends on  $\sqrt{\overline{\Delta v}}$, and its growth is determined by $C_{2}$ which is highly affected by the width of the barrier. Our results recover the analytical phase difference predicted for a delta potential barrier in the limit of $\sigma\rightarrow0$ and very high incident velocities (see Fig. \ref{fig:phase2} (a) with $\sigma=0.011$ $\mu$m). In this limit, the first term of Eq. (\ref{phase}) tends to zero due to its inverse dependence on $\overline{\Delta v}$, the last term, which depends on the width of the barrier, also tends to zero and only remains the term $C_1$ which tends to $\pi/2$ for high enough incident velocities. Thus, retrieving the delta potential barrier behavior \cite{Holmer_2007,Holmer_20071}. 

\subsubsection{Low incident velocities}
\begin{figure}[htbp]
	\centering
		\includegraphics[width=0.50\textwidth]{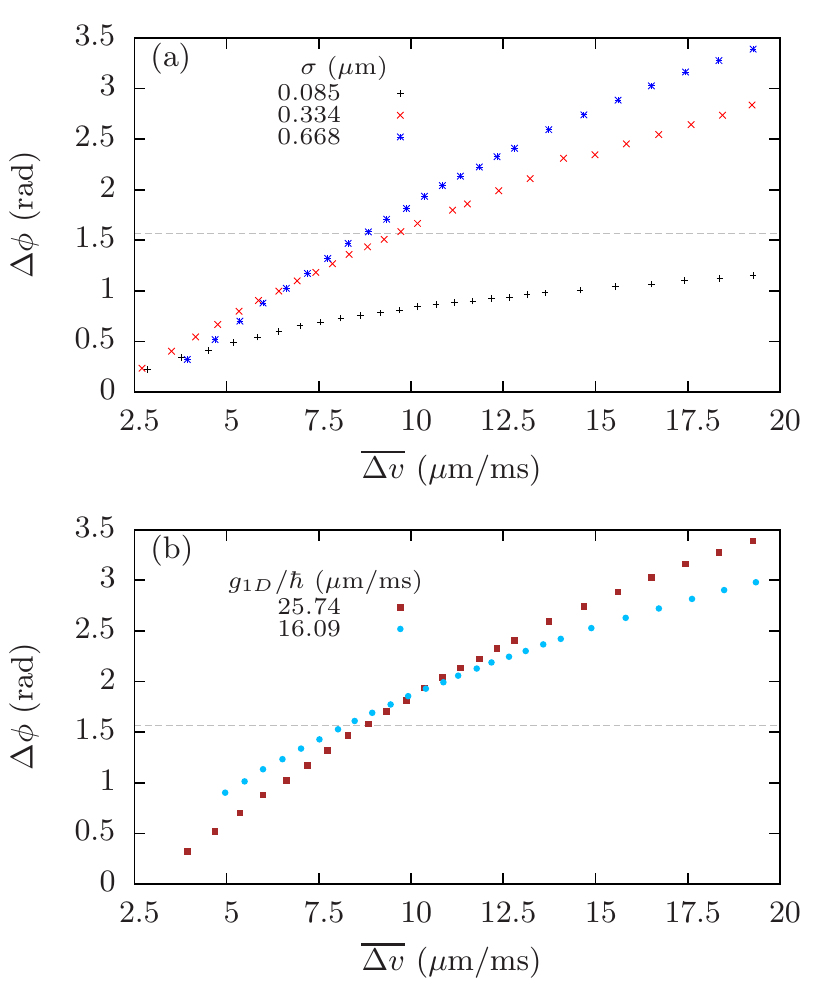}
	\caption{(Color online) Phase difference between the transmitted and reflected solitons as a function of the mean of the absolute values of the velocities of the two split solitons for low velocities of the incident soliton, for a fixed transmission coefficient of $0.5$, and for different values of the width of the barrier and a fixed $g_{1D}/\hbar=25.74$ $\mu$m/ms (a) and for different values of the nonlinear interactions and a fixed width of the barrier $\sigma=0.668$ $\mu$m (b).}
	\label{fig:phase3}
\end{figure}

Here we focus on velocities of the incident soliton and widths of the barrier compatible with the range of parameter values available in current experimental setups \cite{Hulet}. In this regime, we obtain similar general dependences with the width of the barrier and the nonlinearity as the ones described for high incident velocities, but with a considerable deviation with respect to the semi-analytical fit given by Eq. (\ref{phase}). This may be due to the increase of the interaction time between soliton and barrier for low velocities. As for the case of high velocities of the incident soliton, the width of the barrier determines the growth of the phase difference with the mean of the absolute values of the velocities of the two split solitons (Fig. \ref{fig:phase3} (a)), while the nonlinearity affects mainly the arctangent dependence of the phase difference as shown in Fig. \ref{fig:phase3} (b). From Fig. \ref{fig:phase3} (b) we can also notice that the range of represented points is shifted to higher values of the mean of the absolute values of the split solitons as the nonlinearity decreases. This is related with the slowing down of the split solitons (discussed in section \ref{Velocity of the split solitons}) for large nonlinearities. Thus, the same velocity of the incident soliton, $v_0$, leads to different $\overline{\Delta v}$ for different nonlinearities.

	\begin{figure}[htbp]
	\centering
		\includegraphics[width=0.50\textwidth]{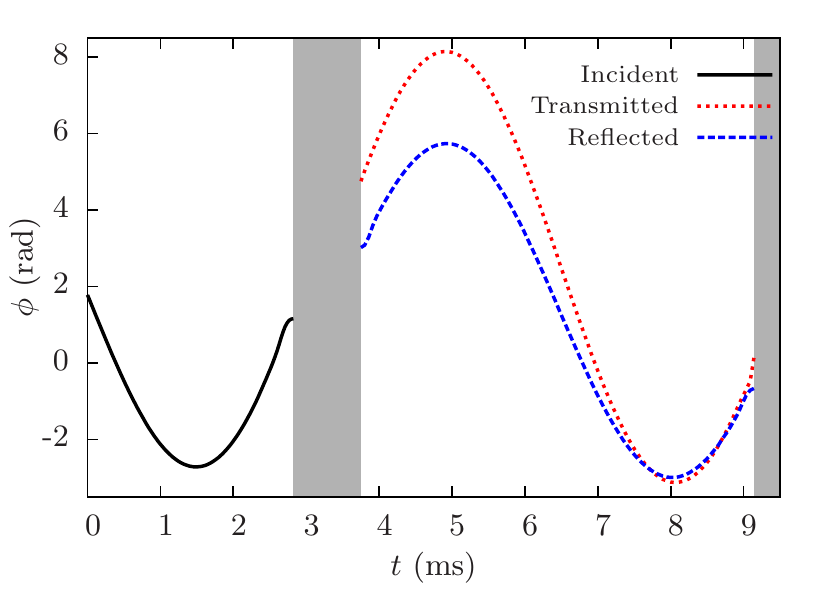}
	\caption{(Color online) Phase evolution of a matter wave bright soliton of $^{7}$Li with $4\times10^3$ atoms (solid line) and the two split solitons (dashed and dotted lines) as a function of time during the complete interferometric sequence. The parameter values are: $V_{b}=16.09\hbar\omega_z$, $\omega_{z}=2\pi\times78$ Hz, $\sigma=0.67$ $\mu$m, $\omega_r=2 \pi\times800$ Hz, $d=-20$ $\mu$m and $a_s=-0.16$ nm }
	\label{fig:phase_evolution}
\end{figure}
\section{Recombination}
\label{sec:Recombination}
\begin{figure*}[htbp]
	\centering
		\includegraphics[width=1.0\textwidth]{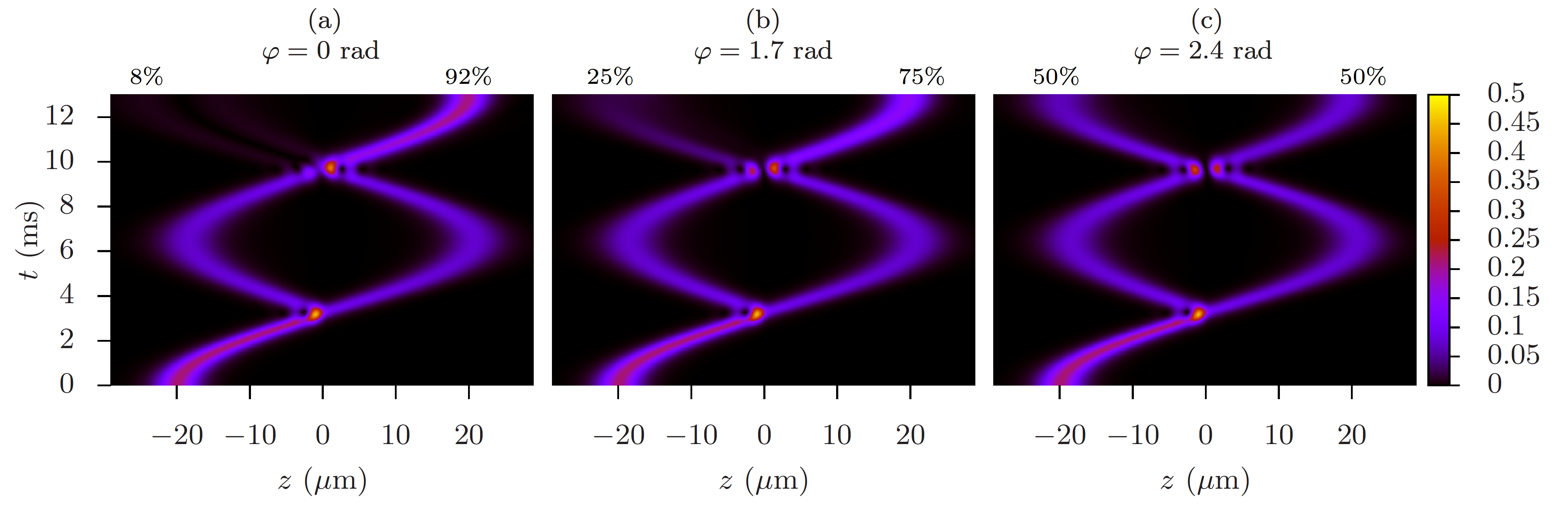}
	\caption{(Color online) Contour plot of the atomic density, $|\Psi(z,t)|^{2}$, during the splitting and recombination processes for an incident soliton of $^{7}$Li with $4\times10^3$ atoms and $a_s=-0.16$ nm centered at $d=-20$ $\mu$m in a harmonic external potential with $\omega_z=2 \pi\times78$ Hz, $\omega_r=2 \pi\times800$ Hz and a Rosen-Morse barrier with height $V_{b}=16.09\hbar\omega_{z}$ and width $\sigma=0.67$ $\mu$m. We plot three different imprinted phases (a) $\varphi=0$ rad, (b) $\varphi=1.7$ rad, (c) $\varphi=2.4$ rad, which produce 92\%, 75\% and 50\% of the initial number of atoms at the right output, respectively.}
	\label{fig:recombination}
\end{figure*}
In this section we study the complete evolution of the interferometric sequence described in Fig. \ref{fig:scheme}. Fig. \ref{fig:phase_evolution} shows the phase evolution of the incident soliton (solid line) and of the two equal-sized split solitons (dashed and dotted lines) as a function of time during the interferometric sequence: (i) the incident soliton (solid line) moves towards the potential barrier, during the first $2.8$ ms of the evolution, converting its potential energy into kinetic energy; (ii) the soliton collides with the Rosen--Morse barrier (gray area between $2.8$ and $3.7$ ms) and splits into two solitons with different phases and different velocities (dashed and dotted lines); (iii) the two solitons perform a dipole oscillation in the trap for approximately $5.4$ ms which provides an oscillatory behavior of the phase, different for each of the solitons due to their different velocities after the splitting (see section \ref{Velocity of the split solitons}); (iv) finally, the two solitons recombine at the position of the barrier (gray area starting at $9.1$ ms) giving two output solitons. The number of atoms at the two outputs of the interferometer depends on the phase difference between the transmitted and reflected solitons at the position of the barrier during the recombination process. We have tested the implementation of the interferometer by means of the phase imprinting method \cite{Dobrek_1999}, modifying instantaneously the phase of one of the arms of the interferometer. Fig. \ref{fig:recombination} (a) shows the density evolution of the incident and split solitons without any imprinted phase. Fig. \ref{fig:recombination} (b) and (c) correspond to the cases of imprinted phases of $\varphi= 1.7$ rad and $\varphi= 2.4$ rad, respectively. Clearly, the number of atoms at the outputs of the interferometer is dominated by the phase difference between the two solitons in the recombination stage. 


\section{Conclusions}
\label{conclusions}

We have studied a matter wave bright soliton interferometer composed of a harmonic external potential trap with a Rosen--Morse potential barrier at its center. We have focused on the analysis of the splitting process for the case where the two split solitons have exactly the same number of atoms. First, we have shown that the area of the Rosen-Morse barrier necessary to obtain the equal-sized splitting retrieves the delta behavior for very thin barriers and very high incident velocities. Otherwise, a quadratic behavior of the area of the barrier with the velocity of the incident soliton appears. We have also reported that the velocities of the reflected and transmitted solitons are strongly affected by the nonlinearity being both solitons slowed down, and eventually trapped at the position of the barrier, for high enough nonlinearities. In addition, we have found that, in general, the reflected soliton is slower than the transmitted one. We have also characterized the phase difference between the two split solitons. For high velocities of the incident soliton we have derived a semi-analytical fit that reproduces the main dependences on the velocity, width of the barrier and nonlinearity for the equal-sized splitting. We have also recovered the delta behavior in the limit of high incident velocities and extremely thin barriers. Finally, we have analyzed the recombination process, studying first the phase evolution in the full interferometric sequence, and then we have tested the performance of the interferometer, introducing a phase difference between its two arms by means of the phase imprinting method showing that the number of atoms at each of the outputs is strongly affected by the introduced phase difference.



\begin{acknowledgments}
We thank Albert Benseny, Juan Luis Rubio, and Daniel Viscor for fruitful discussions and comments. We acknowledge support from the Spanish Ministry of Economy and Competitiveness under contract FIS2011-23719 and from the Catalan Government under contract SGR2009-00347. J. Polo also acknowledges financial support from the FPI grant with reference BES-2012-053447.
\end{acknowledgments}

\appendix
\section{Matter wave bright solitons solutions}

The bright soliton solution of the homogeneous one dimensional GPE with attractive interatomic interactions can be expressed as \cite{Pethick_2001}:
\begin{equation}
\label{soliton}
\Psi(z,t)=\sqrt{\frac{\alpha}{2}}\sech\left[\alpha(z-z_0 -v t)\right] e^{i\phi(z,t)}
\end{equation}
where:
\begin{equation}
\label{phase_soliton}
\phi(z,t)=\frac{m v}{\hbar}\left((z-z_0)-\frac{v t}{2}\right)+\frac{\hbar}{2m}\alpha^{2} t+\theta,
\end{equation}
$v$ is the soliton velocity, $z_0$ is its initial position, $\theta$ is an arbitrary initial phase and $\alpha=N\omega_{r}a_{s}m/\hbar$ with $N$, $\omega_{r}$, $a_{s}$ and $m$ corresponding to the atom number, frequency of the radial confinement, s-wave scattering length and atomic mass, respectively. Even though Eq. (\ref{soliton}) is strictly valid only in the homogeneous case, the ground state of an attractive BEC in a quasi-one dimensional geometry with sufficiently weak axial confinement takes the form of the bright soliton solution of the homogeneous GPE \cite{Billam_2012}.

In the absence of external potential, the total energy of the soliton can be separated into three contributions \cite{Ernst_2010}:
\begin{equation}
E_t=E^{p}_{k}+E^{v}_{k}+E_{int}
\end{equation}
where:
\begin{equation}
E^{p}_{k}=\frac{\hbar^{2}}{2m}\int{dz \left|\frac{\partial|\Psi(z)|}{\partial z}\right|^{2}}=\frac{\alpha^{2}\hbar^{2}}{6 m},
\end{equation}
\begin{equation}
E^{v}_{k}=\frac{\hbar^{2}}{2m}\int{dz \left|\left|\Psi(z)\right|\frac{\partial}{\partial z}\exp{[i \varphi(z,t)]}\right|^{2}}=\frac{m}{2}v^{2},
\end{equation}
\begin{equation}
E_{int}=-\frac{g_{1D}}{2}\int{dz \left|\Psi(z)\right|^{4}}=-\frac{\alpha^{2}\hbar^{2}}{3 m},
\end{equation}
being $E^{p}_{k}$ the quantum pressure term, $E^{v}_{k}$ the kinetic energy given by the gradient of the phase of the soliton, and $E_{int}$ the energy due to the nonlinearity.

The general $n$-soliton solution of the homogeneous GPE can be written, for large enough spatial separation among them, as the sum of $n$ single soliton solutions \cite{Gordon_1983}:
\begin{equation}
\label{sum_sol}
\Psi(z,t)=\sum_{j=1}^n\Psi_{j}(z,t)
\end{equation}
with:
\begin{equation}
\label{solitoni}
\Psi_{j}(z,t)=A_{j}\sqrt{\frac{\alpha}{2}}\sech\left[A_{j}\alpha(z-z_{0 j} -v_{j} t)+q_j\right]
\end{equation}
$$
\times\exp{\left[i\left\{\frac{m v_{j}}{\hbar}\left((z-z_{0 j})-\frac{v_{j} t}{2}\right)+\frac{\hbar}{2m}A_{j}^{2}\alpha^{2} t+\theta_{j}+\psi_{j}\right\}\right]},
$$
where $A_{j}$ corresponds to the amplitude of the $j$th soliton, $q_j$ and $\psi_j$ are the position and phase shift, respectively, which are given by \cite{Gordon_1983}:
\begin{equation}
\label{phaseshift}
q_j +i \psi_j=\sum_{k\neq j}{\pm \ln{\left[\frac{A_j +A_k +i(v_{j}-v_{k})m/2\hbar\alpha}{A_j -A_k +i(v_{j}-v_{k})m /2\hbar\alpha}\right]}}
\end{equation}

\bibliographystyle{apsrev4-1}

\bibliography{shorttitles,references}

\end{document}